\documentclass[amsmath,amssymb,numbers]{revtex4}

\begin{document}
\bibliographystyle{plainnat}
\title{Finite Geometry and the Radon Transform}

\author{M. Revzen}
\affiliation {Department of Physics, Technion - Israel Institute of Technology, Haifa
32000, Israel}

\date{\today}

\begin{abstract}
Finite geometry is used to underpin operators acting in finite, d, dimensional Hilbert space.
Quasi distribution and Radon transform underpinned with finite dual affine plane geometry
(DAPG) are defined in close analogy with the continuous $(d\rightarrow \infty)$ Hilbert space case.
An essential role in these definitions play projectors of states of mutual unbiased bases (MUB) and their
Wigner function-like mapping onto the generalized phase space that lines and points of DAPG constitutes.
\end{abstract}


\maketitle

\section {  Introduction}

In this article we consider, in a d-dimensional Hilbert space, the underpinning of
information theoretic related operators with finite geometry.\\

The general idea of such a relation was studied intensely,
\cite{wootters4,saniga,planat1,planat2,combescure}. Our approach is focused on a particular
branch of finite geometrical system, \cite{wootters1,wootters2,wootters3,tomer}: dual
affine plane geometry (DAPG) whose main features are outlined in  section III. The
operators that we discuss are the operators involved in "Mutual Unbiased Bases" (MUB)
\cite{wootters1,wootters2,bengtsson,amir}. The study is confined to dimensionality, d=p, a
prime, $\ne 2$. We itemize the relevant definitions/properties concerning these operators
in the next section, section II, in our attempt to have this paper
self contained.\\
The general idea of geometrically underpinned Hilbert space is as follows. A Geometry is
defined by interrelation among points and lines. Points are considered the fundamental
entities and lines are aggregates of points. We specify the rules for the so called
(finite) affine plane geometry (APG) and its dual (DAPG) in section III. A specific
arrangement of points and lines abiding by such rules form a realization of the geometry.
The association of Hilbert space operators with geometrical points and lines define a
geometrical underpinning of the Hilbert space operators. Consistency of an underpinning is
attested to by the consistency of the operators implied physical results/relation and the
interrelations dictated by the geometry. The present study relates to the so called state
reconstruction \cite{wootters3,pier}: the Hilbert space MUB state projectors allow the {\it
diagonal} elements (of all MUB states) of the density operator to define in full the
density operator. We illustrate this with a study of  Radon transform \cite{radon,yves}
which, hitherto, was confined to studies in continuum. Consistency may here be checked by
state reconstruction
via the inverse transform.\\

The focal issue in the present analysis is finite dimensional Radon transform. This
transform, for the continuous case ($d\rightarrow \infty$) \cite{radon,yves}, is widely
used both in down to earth application \cite{cho,guy,freeman} mathematical studies
\cite{halgason,deans}, and state reconstruction analysis \cite{ulf,schleich,raymer,faqir}.
In the present work, dealing with finite dimensional Radon transform,  it is, much like in
the continuous case, analyzed via quasi distribution, viz Wigner-like function defined here
in the finite phase space-like points and lines of DAPG. \\

 The Radon transformation \cite{radon,li,pier} involves angular variables and thus its
 formulation in finite
(Hilbert) space dimensionality is somewhat intricate. We overcome this by adopting a
"physical" approach whose rationale, within the continuum, $d\rightarrow \infty$ case is as follows.
The Wigner function, $W_{A}(q,p)$ maps an operator,
 $\hat{A}$ in Hilbert space onto a c-number function in phase space \cite{moyal,ellinas}. When the
 operator is the state density operator, $\hat{\rho}$, the resultant Wigner function, $W_{\rho}(q,p),$
is quasi distribution, i.e. it enjoys many attributes of a distribution (it may, however, become negative)
\cite{wigner,ulf,schleich,walls}. A particularly attractive attribute that underscores its role as
quasi distribution is its marginal with respect to the position q. Thus integrating along a vertical
line (the p coordinate) for a fixed q=x':
 $\int\frac{dqdp}{2\pi}\delta(x'-q) W_{\rho}(q,p),$
 gives the probability for the system in a state $\rho$ (i.e. whose quasi distribution is $W_{\rho}(q,p)$)
 to have its position value x'. In other words this integral equals the expectation value of
 the  projector $|x'\rangle \langle x'|$ that projects the eigen state, $|x'\rangle$, of  the operator
 $\hat{x}.$ The Wigner phase space mapping of this projector is $W_{|x'><x'|}(q,p)=\delta(x'-q)$. This
 continues to hold when we replace the
position operator by arbitrary MUB state projector, $|x',\theta\rangle \langle \theta,x'|$,
with $|x',\theta\rangle$ the eigenfunction of $\hat{X}_{\theta}= \hat{x}cos\theta
+\hat{p}sin\theta$, eigen value x' \cite{schleich,ulf,walls}. In this case, since the
Wigner function $W_{|x',\theta><\theta,x'|}(q,p)=\delta(x'-qcos\theta-psin\theta),$ the
integral is along the line $y=-qsin\theta + pcos\theta$. Thus here too the marginal relates
to MUB projectors \cite{wootters4,ent,pier} and we have for the probability,
$P(x',\theta:\rho)$, of the state $\rho$ being found in $x'\theta\rangle\langle\theta,x'|$
($C=cos\theta,\;S=sin\theta$),
\begin{equation}
P(x',\theta:\rho)\equiv tr\big(\rho |x',\theta\rangle \langle\theta,x'|\big)=
\int\frac{dqdp}{2\pi}\delta(x'-Cq-Sp)W_{\rho}(q,p).
\end{equation}
In this form we recognize the marginal probability of the MUB projector's expectation value
as the Radon transform of the quasi distribution, $W_{\rho}(q,p)$. An explicit account is
given in section II. We use these observations as our guide for the definition of the
finite dimensional Radon transform. Indeed the inversion of the transform is, in effect, a
state reconstruction: it gives the state's quasi distribution,  in terms of
the probabilities $P(x',\theta:\rho)$.\\

In the finite dimensional (d=p, a prime $\ne 2$) Hilbert space we define a Wigner
function-like mappings of operators onto lines and points of DAPG. We show that our Wigner
function-like mapping of the density operator, $V_{\rho}(j),$ onto DAPG lines, $L_j,$ - has
all the quasi distribution attributes possessed by $W_{\rho}(q,p)$ in the $d \rightarrow
\infty$ case. The marginals of $V_{\rho}(j)$ that give the probabilities of the system
being in some eigen function of the finite dimensional MUB sets are now given as summation
are along the DAPG line, $L_j$ these being  assured via the function $\Lambda_{\alpha, j},$
(defined in the text, $\alpha$  designates a DAPG point and j a line.) which corresponds to
the delta function in the $d\rightarrow \infty.$ Moreover, the marginals here, much like
the continuous case, are informationally complete i.e. they allow the reconstruction of the
corresponding (quasi) distribution which, in turn, \cite{schleich,pier} determine the
state. Thus our finite dimensional Radon transform is the marginals of the quasi
distributions pertaining to the
projectors of MUB states in complete analogy with the $d \rightarrow \infty$ case.\\

 The inversibility attribute
of the transform allows the extraction of the system's state from lines' summations that
is informationally complete. For clarity we give a brief explicit review of this for the
continuous case in section II. The finite geometry  with which we underpin the theory, viz
dual affine plane geometry (DAPG) is outlined in section III. MUB is defined in section IV.
Section V contains the basic underpinning theory. Here and in the succeeding section we
list some useful formulae. The formulation of the finite dimension Radon transform with its
explicit inversion as well as the definition our quasi distribution are contained in the
succeeding section, section VII. Section VIII introduces underpinning of finite dimensional Hilbert space operators with affine plane geometry, APG. In the last section, section IX, we summarize and
discuss the results.\\

\section{   Mutually Unbiased Bases and Radon Transform - the $d \rightarrow \infty$ case.}

The rationale of our analysis is the relation between informationally complete measurements
which we enumerate via mutual unbiased bases (MUB) and the sought after state of the
system. In this section we outline the (known e.g. \cite{schleich}) approach for the
continuous case thereby clarifying, we hope, the (known, \cite{ulf,schleich}) result that
the state, actually its Wigner representative function, is determined from the {\it
diagonal} elements of the density operator with respect to (all) the states in all the MUB
bases. To this end we briefly review the notion of MUB their measurements and the
extraction thereof the Wigner function of
the density operator.\\

Mutual unbiased bases, MUB, in concept were introduced by Schwinger \cite{schwinger} in his
studies of vectorial bases for Hilbert spaces that exhibit ``maximal degree of
incompatibility''. The  eigenvectors of $\hat{x}$  and $\hat{p}$, $|x\rangle$ and
$|p\rangle$ respectively are example of such bases. The information theoretical oriented
appellation ``mutual unbiased bases'' were introduced by Wootters \cite{wootters2}. Two
complete,
  orthonormal vectorial bases, ${\cal B}_1,\;{\cal B}_2$,
 are said to be MUB if and only if (${\cal B}_1\ne {\cal B}_2)$
\begin{equation}
\forall |u\rangle,\;|v \rangle\; \epsilon \;{\cal B}_1,\;{\cal B}_2 \;resp.,\;\;|\langle
u|v\rangle|=const.
\end{equation}
i.e. the absolute value of the scalar product of vectors from {\it different} bases is
independent of the vectorial label within either basis. This implies that if a state vector
is measured to be in one of the states,
 $|u\rangle,$ of ${\cal B}_1$ it is equally likely to be in any of the states $|v\rangle$
of any other  MUB,  ${\cal B}_2$. (The value of the $|\langle u|v\rangle|$ may depend on
the {\it bases}, ${\cal B}_1,{\cal B}_2$, which indeed is the case for the limit  $d
\rightarrow \infty$, the continuous dimensionality.) MUB are found to be of interest in
several fields. The ideas are useful in a variety of
 cryptographic protocols \cite{ekert} and signal analysis \cite{vourdas}.\\

We now outline, for the continuous Hilbert space dimensionality,  some salient MUB features
\cite{amir,mello1,pier,ent}. Consider the so called quadrature \cite{schleich,ulf,walls} operator
$\hat{X}_{\theta}$ and its eigen states state $|x',\theta \rangle,$
\begin{equation}\label{x,theta}
\hat{X}_{\theta}|x',\theta\rangle=x'|x',\theta \rangle.
\end{equation}
 We now show that the states, $|x',\theta \rangle$ form a complete orthonormal basis for the
 (rigged) Hilbert space,
 with the vectors labeled by x' and the basis by
$\theta$,  and bases with different $\theta,\;\;0\le \theta\le \pi$, are MUB.  We begin by
noting, that, as can be easily checked \cite{schleich,ulf, amir}
\begin{equation}\label{x}
\hat{X}_{\theta}=U^{\dagger}(\theta)\hat{x}U(\theta)=C\hat{x}+S\hat{p},
\end{equation}
where
\begin{equation}
U(\theta)=e^{-i\theta \hat{a}^{\dagger} \hat{a}};\;\;\hat{a}=
\frac{1}{\sqrt2}(\hat{x}+i\hat{p}),
\end{equation}
and $C=Cos\theta,\;S=Sin\theta$. Thus the solution to Eq(\ref{x,theta}) may be written in
terms of the eigenvectors of the position operator \cite{mello1,mello2,amir,pier,ent},
\begin{equation}\label{Un}
|x,\theta\rangle=U^{\dagger}(\theta)|x \rangle.
\end{equation}
This defines our phase choice \cite{oded}. We now use the well known result for the
harmonic oscillator propagator,  \cite{pier,larry}, to get the  x representation solution,
\begin{equation} \label{overlap}
\langle x'|x;\theta\rangle = \frac{1}{\sqrt{-i2\pi S}}e^{-\frac{1}{2S}([x^2+x'^2]C-2xx')}.
\end{equation}

This form  ascertains \cite{pier} that $\lim_{\theta\rightarrow 0}\langle
x'|x;\theta\rangle =\delta(x-x')$, and $\lim_{\theta\rightarrow \pi/2}\langle
x'|x;\theta\rangle=e^{xx'}/\sqrt{2\pi}.$ This phase differs from the more common one
\cite{wootters1} and leads to an expression that is symmetric $x \leftrightarrow x'$, a
property that facilitates several calculations. We now verify, by direct calculation, that
the bases $\{|x;\theta\rangle \}$ and $\{|x';\theta'\rangle\}$
 with $\theta \ne \theta'$  are MUB:
\begin{equation}
|\langle x';\theta'|x;\theta \rangle|=|\langle x'|U^{\dagger}(\theta-\theta')|x
\rangle|=\frac{1}{\sqrt{2\pi|S(\theta,\theta')}|},
\end{equation}
 $S(\theta,\theta')=Sin(\theta-\theta')$. Thus the number $|\langle x';\theta'|x;\theta\rangle|$
 is {\it
in}dependent of the vectorial labels x, x'. We used  the relation
$U(\theta')U^{\dagger}(\theta)=U^{\dagger}(\theta-\theta')$ and Eq(\ref{Un}).\\
The Wigner function that represents an arbitrary operator $\hat{A}$  in phase space
 \cite{wigner,schleich,ulf,walls,moyal,ellinas} is given by,
\begin{equation} \label{w1}
W_{A}(q,p)\;=\;\int dye^{ipy}\langle q-y/2|\hat{A}|q+y/2 \rangle.
\end{equation}
Thus the Wigner function maps an Hilbert space operator to a c-number function in phase space.
An important attribute of this mapping is
\begin{equation}\label{w}
tr\hat{A}\hat{B}= \int \frac{dq dp}{2\pi}W_{A}(q,p)W_{B}(q,p).
\end{equation}
It can be shown \cite{schleich,ulf} that the Wigner function of the density operator is real
though {\it not} non negative (in general) this, with Eq.{\ref{w}), instigates its
classification as a {\it quasi} distribution in phase space. This completes our brief
 review of MUB and quasi distributions for  continuous
Hilbert space \cite{amir}. Now we turn to the state reconstruction and Radon transform role
for this case.\\
The probability of obtaining x' upon measuring the quadrature operator
 $\hat{X}_{\theta}$, Eq.(\ref{x}) for the  state $\rho$ is, \cite{ulf,schleich,pier}
 \begin{equation}\label{g}
 \rho(\theta,x')\equiv tr(\rho |x';\theta \rangle \langle \theta,x'|)=\int
 \frac{dqdp}{2\pi}W_{\rho}(q,p)W_{|x',\theta\rangle}(q,p),
 \end{equation}
where we used Eq.(\ref{w}) to write the RHS of the equation. Explicit evaluating
$W_{X_{\theta}}(q,p),$ via Eq.(\ref{w1}), gives for $\rho(\theta,x')$,
\begin{equation}\label{r}
\rho(\theta,x')=\int\frac{dqdp}{2\pi}W_{\rho}(q,p)\delta(x'-Cq-Sp).
\end{equation}
This identifies \cite{freeman} $\rho(\theta,x')$ as the Radon transform of $W_{\rho}(q,p)$.
Thus the Radon transform \cite{yves,li} ${\cal{R}}f(L)$  of a function f(x,y) is the integral
of the function over straight line, L. The line in Eq.(\ref{r}) is a line in phase space
whose given by $x'=Cq+Sp$. (The full transform requires the integration over all parallel
lines i.e. over all values of x'). Solving for $W_{\rho}(q,p)$ in terms of the observables
$\rho(\theta,x')$ \cite{yves}:
\begin{equation}\label{y}
W_{\rho}(q,p)={\cal{R}}^{- 1}({\cal{R}}W_{\rho})(q,p)={\cal{R}}^{- 1}\rho(\theta,x').
\end{equation}
Here ${\cal{R}}$ represents Radon transform. Direct calculation gives \cite{ulf, schleich,
amir, pier}
\begin{equation}
W_{\rho}(q,p)=-\frac{1}{\pi}\int_0^{\pi}d\theta {\cal{P}}\int_{-\infty}^{\infty} dx'
\frac{\partial{\rho(x,\theta}/\partial{x}}{x-q C-p S}.
\end{equation}
Thus inverting the Radon transformation is state reconstruction as the state $\rho$ is
determined once its Wigner representative is known \cite{schleich}.\\
Note that $\rho(\theta,x')$ plays a double role: (a) As a marginal it is required via
Eq.(\ref{w}), that the RHS of Eq.(\ref{g}) contain the Wigner function of the MUB state
projector, $|x',\theta\rangle \langle \theta,x'|.$ Alternatively, (b), as the Radon
transform of $W_{\rho}(q,p)$, this is consistent because the Wigner function of the
projector is a delta function
assuring, in the integration, that x' equals Cq+Sp in phase space.\\

 \section{   Finite Geometry and Hilbert Space Operators}

We now briefly review the essential features of finite geometry required for our study
\cite{grassl,diniz,shirakova,tomer,wootters4}.\\
A finite plane geometry is a system possessing a finite number of points and lines. There
are two kinds of finite plane geometry: affine and projective. We shall confine ourselves
to affine plane geometry (APG) which is defined as follows. An APG is a non empty set
whose
elements are called points. These are grouped in subsets called lines subject to:\\
1. Given any two distinct points there is exactly one line containing both.\\
2. Given a line L and a point S not in L ($S \ni L$), there exists exactly one line L'
containing S
such that $L \bigcap L'=\varnothing$. This is the parallel postulate.\\
3. There are 3 points that are not collinear.\\
It can be shown \cite{diniz,shirakova} that for $d=p^m$ (a power of prime) APG can be
constructed (our study here is for d=p) and the following properties are, necessarily,
built in: \\
a. The number of points is $d^2;$ $S_{\alpha},\;{\alpha = 1,2,...d^2}$ and the number of
lines is d(d+1); $L_j,\;j=1,2....d(d+1)$.\\
b. A pair of lines may have at most one point in common: $L_j\bigcap
L_k=\lambda;\;\lambda=0,1\;for j \ne k$.\\
c. Each line is made of d points and each point is common to d+1 lines:
$L_j=\bigcup_{\alpha}^d S_{\alpha}^j$, $S_{\alpha}=\bigcap_{j=1}^{d+1}L_j^{\alpha}.$\\
d. If a line $L_j$ is parallel to the distinct lines $L_k\;and\;L_i$ then $L_k \parallel
L_i$. The $d^2$ points are grouped in sets of d parallel lines. There are d+1 such
groupings.\\
e. Each line in a set of parallel lines intersect each line of any other set:
$L_j\bigcap L_k=1;\;L_j \nparallel L_k.$ \\
The above items will be referred to by APG (x), with x=a,b,c,d or e.\\

The existence of APG implies \cite{diniz,grassl,shirakova}the existence of its dual
geometry DAPG wherein the points and lines are interchanged. Since we shall study
extensivebly this, DAPG, we list the
corresponding properties for it. We shall refer to these by DAPG(y):\\
a. The number of lines is $d^2$, $L_j,\;j=1,2....d^2.$ The number of points is d(d+1),
$S_{\alpha},\;{\alpha = 1,2,...d(d+1)}.$\\
b. A pair of points on a line determine a line uniquely. Two (distinct) lines share one and only
one point.\\
c. Each point is common to d lines. Each line contain d+1 points.\\
d. The d(d+1) points may be grouped in sets of d points no two of a set
share a line. Such a set is designated by $\alpha' \in \{\alpha \cup M_{\alpha}\},\;
\alpha'=1,2,...d$. ($M_{\alpha}$ contain all the points not connected to $\alpha$ - they
are not connected among themselves.) i.e. such a set contain d disjoined (among themselves)
points. There are d+1 such sets:
\begin{eqnarray}
\bigcup_{\alpha=1}^{d(d+1)}S_{\alpha}&=&\bigcup_{\alpha=1}^d R_{\alpha};\nonumber\\
R_{\alpha}&=&\bigcup_{\alpha'\epsilon\alpha\cup M_{\alpha}}S_{\alpha'}; \nonumber \\
R_{\alpha}\bigcap R_{\alpha'}&=&\varnothing,\;\alpha\ne\alpha'.\nonumber
\end{eqnarray}
e. Each point of a set of disjoint points is connected to every other point not in its
set.\\

A particular arrangement of lines and points that satisfies APG(x), x=a,b,c,d,e is referred to as a realization of APG. Similar prescription holds for DAPG.

We now consider a particular realization of DAPG of dimensionality $d=p,\ne 2$ which is the
basis of our present study. We arrange the aggregate the d(d+1) points, $\alpha$, in a
$d\cdot(d+1)$matrix like rectangular array of d rows and d+1 columns. Each column is made
of a set of d points  $R_{\alpha}=\bigcup_{\alpha'\epsilon\alpha\cup
M_{\alpha}}S_{\alpha'};$  DAPG(d). We label the columns by b=-1,0,1,2,....,d-1 and the rows
by m=0,1,2...d-1.( Note that the first column label of -1 is for convenience and does not
designate negative value of a number.)  Thus $\alpha=m(b)$ designate a point by its row, m,
and its column, b; when b is allowed to vary - it designate the point's row position in
every column. We label the left most column by b=-1 and with increasing values of b, the
basis label, as we move to the right. Thus the right most column is b=d-1. We now assert
that the d+1 points, $m_j(b), b=0,1,2,...d-1,$  and  $m_j(-1)$, that form the line j which
contain the two (specific) points m(-1) and m(0) is given by (we forfeit the subscript j -
it is implicit),
\begin{eqnarray}\label{m(b)}
m(b)&=&\frac{b}{2}(c-1)+m(0),\;mod[d]\;\;b\ne -1, \nonumber \\
m(-1)&=&c/2.
\end{eqnarray}

The rational for this particular form is clarified in Section V. Thus a line j is
parameterized fully by $j=(m(-1),m(0))$. We now prove that the set $j=1,2,3...d^2$ lines
covered by Eq.(\ref{m(b)}) with the points as defined above form a DAPG.\\
\noindent 1. Since each of the  parameters, m(-1) and m(0), can have d values the number of
lines $d^2$; the number of points in a line is evidently d+1.  DAPG(a).\\
\noindent 2. The linearity of the equation precludes having two points with a common value
of b on the same line. Now consider two points on a given line, $m(b_1),m(b_2);\;b_1\ne
b_2$. We have from Eq.(\ref{m(b)}), ($b\ne -1,\;b_1 \ne b_2$)
\begin{eqnarray}\label{twopoints}
m(b_1)&=&\frac{b_1}{2}(c-1)+m(0),\;\;mod[d]\nonumber\\
m(b_2)&=&\frac{b_2}{2}(c-1)+m(0),\;\;mod[d].
\end{eqnarray}
These two equation determine uniquely ({\it for d=p, prime}) m(-1) and m(0). DAPG(b).\\
\noindent For fixed point, m(b), $c\Leftrightarrow m(0)$ i.e the number of free parameters
is d (the number of points on a fixed column). Thus each point is common to d lines. That
the line contain d+1 is obvious. DAPG(c).\\
\noindent 3. As is argued in 2 above no line contain two points in the same column (i.e.
with equal b). Thus the d points, $\alpha,$ in a column form a set
$R_{\alpha}=\bigcup_{\alpha'\epsilon\alpha\cup M_{\alpha}}S_{\alpha'},$ with trivially
$R_{\alpha}\bigcap R_{\alpha'}=\varnothing,\;\alpha\ne\alpha',$ and
$\bigcup_{\alpha=1}^{d(d+1)}S_{\alpha}=\bigcup_{\alpha=1}^d R_{\alpha}.$ DAPG(d).\\
\noindent 4. Consider two arbitrary points {\it not} in the same set, $R_{\alpha}$ defined
above: $m(b_1),\;m(b_2)\;\;(b_1\ne b_2).$ The argument of 2 above states that, {\it for
d=p}, there is a unique solution for the two parameters that specify the line containing
these points. DAPG(e).\\
We illustrate the above for d=3, where we explicitly specify the points contained in the
line $j=\big(m(-1)=(1,-1),m(0)=(2,0)\big)$
\[ \left( \begin{array}{ccccc}
m\backslash b&-1&0&1&2 \\
0&\cdot&\cdot&\cdot&(0,2)\\
1&(1,-1)&\cdot&(1,1)&\cdot\\
2&\cdot&(2,0)&\cdot&\cdot\end{array} \right)\].\\
For example the point m(1) is gotten from
$$ m(1)= \frac{1}{2}(2-1)+2=1\;\;mod[3]\;\;\rightarrow\;m(1)=(1,1).$$

\section{   Finite dimensional Mutual Unbiased Bases, MUB, Brief Review}

In a finite, d-dimensional, Hilbert space two complete, orthonormal vectorial bases, ${\cal
B}_1,\;{\cal B}_2$,
 are said to be MUB if and only if (${\cal B}_1\ne {\cal B}_2)$

\begin{equation}
\forall |u\rangle,\;|v \rangle\; \epsilon \;{\cal B}_1,\;{\cal B}_2 \;resp.,\;\;|\langle
u|v\rangle|=1/\sqrt{d}.
\end{equation}
The physical meaning of this is that knowledge that a system is in a particular state in
one basis implies complete ignorance of its state in the other basis.\\
Ivanovic \cite{ivanovich} proved that there are at most d+1 MUB, pairwise, in a
d-dimensional Hilbert space and gave an explicit formulae for the d+1 bases in the case of
d=p (prime number). Wootters and Fields \cite{wootters2} constructed such d+1 bases for
$d=p^m$ with m an integer. Variety of methods for construction of the d+1 bases for $d=p^m$
are now available
\cite{tal,klimov2,vourdas}. Our present study is confined to $d=p\;\ne2$.\\
 We now give explicitly the MUB states in conjunction with the algebraically complete
 operators \cite{schwinger,amir} set:
 $\hat{Z},\hat{X}$.  Thus we label the d distinct states spanning the Hilbert space,
 termed
 the computational basis, by $|n\rangle,\;\;n=0,1,..d-1; |n+d\rangle=|n\rangle$
\begin{equation}
\hat{Z}|n\rangle=\omega^{n}|n\rangle;\;\hat{X}|n\rangle=|n+1\rangle,\;\omega=e^{i2\pi/d}.
\end{equation}
The d states in each of the d+1 MUB bases \cite{tal,amir}are the states of computational basis and
\begin{equation} \label{mxel}
|m;b\rangle=\frac{1}{\sqrt
d}\sum_0^{d-1}\omega^{\frac{b}{2}n(n-1)-nm}|n\rangle;\;\;b,m=0,1,..d-1.
\end{equation}
Here the d sets labeled by b are the bases and the m labels the states within a basis. We
have \cite{tal}
\begin{equation}\label{tal1}
\hat{X}\hat{Z}^b|m;b\rangle=\omega^m|m;b\rangle.
\end{equation}
For later reference we shall refer to the computational basis (CB) by b=-1. Thus the above
gives d+1 bases, b=-1,0,1,...d-1 with the total number of states d(d+1) grouped in d+1 sets
each of d states. We have of course,
\begin{equation}\label{mub}
\langle m;b|m';b\rangle=\delta_{m,m'};\;\;|\langle m;b|m';b'\rangle|=\frac{1}{\sqrt d},
\;\;b\ne b'.
\end{equation}
This completes our discussion of MUB.\\

\section{  Geometric Underpinning of MUB Quantum Operators}

 We now consider a DAPG as underpinning a two sets of operators acting in a
d-dimensional Hilbert space, these are
$$ \hat{A}_{\alpha};\;\;\alpha= 1,2....d(d+1),\;\;\hat{P}_j;\;\; j=1,2,..d^2.$$

Here $\hat{A}_{\alpha}$ are associated  with the d(d+1) points, $S_{\alpha}$ while
$\hat{P}_j$ are associated  with the $d^2$ lines, $L_j$.\\
We now define interrelation among the operators of the sets in a way that reflects the geometry,
\begin{equation}\label{oprel3}
\hat{A}_{\alpha}=\frac{1}{d}\sum_{j\in \alpha}^{d}\hat{P}_j.
\end{equation}
These entail, using DAPG(a,b,d),
\begin{equation}\label{oprel}
\hat{P}_j=\sum_{\alpha \in j}\hat{A}_{\alpha}-\frac{1}{d}\sum^{d^2}_{j'}\hat{P}_{j'}.
\end{equation}

We now list some known finite dimensional features. These will allow its underpinning with
DAPG which are presented in the succeeding Section.\\

We consider d=p, a prime. For d=p we may construct d+1 MUB
\cite{ivanovich,tal,wootters1,wootters4}. Returning to our state labeling we label the MUB
states by $|m,b\rangle.$ We designate the computational basis, CB, by  b=-1,  while
$b=0,1,2...d-1$ labels the eigenfunction of, resp. $XZ^b$. m labels the state within a
basis. (Note that assigning b=-1 to the CB is for reference only.) The projection operator defined by,
\begin{equation}\label{ptop}
\hat{A}_{\alpha}\equiv |m,b\rangle \langle
b,m|;\;\alpha=\{b,m\};\;\;b=-1,0,1,2...d-1;\;m=1,2,..d.
\end{equation}
The point label, $\alpha=(m,b)$ is now associated with the projection operator,
$A_{\alpha}$ . We now consider a realization, possible for d=p, a prime , of a d
dimensional DAPG,  as points marked on a rectangular whose width ( x-axis) is made of d+1
column of points, each column is labelled by b, and its height (y axis) is made of d points
each marked with m. The total number of points is d(d+1) - there are d points in each of
the d+1 columns. We associate the d points $m=1,2,...d$, in each set labelled by b
$S_{\alpha};\;\alpha \sim (m,b)$ to the {\it disjointed} points of DAPG(d), viz. for fixed
b $\alpha' \in \alpha \cup M_{\alpha}$ form a column. The columns are arranged according to
their basis label, b. The first being b=-1, $\alpha_{-1}=(m,-1);m=0,1,...d-1,$ signifies
the computational basis (CB). The lines are now made of d+1 points each of different b. To
a line $L_j$ we associate an operator $\hat{P}_j$. Now DAPG(c) (and Eq.(\ref{oprel3}))
implies:
\begin{eqnarray}
\sum_m^d |m,b\rangle\langle b,m|&=&\sum_{\alpha' \in \alpha\cup M_{\alpha}}^d\hat{A}_{\alpha'}
=\hat{I} \nonumber \\
\sum_{\alpha}^{d(d+1)}\hat{A}_{\alpha}&=&(d+1)\hat{I}.
\end{eqnarray}
Returning to Eq.(\ref{oprel}) we have,
\begin{eqnarray}\label{mub2}
\sum_j^{d^2}\hat{P}_j&=&d\hat{I}.\nonumber \\
\hat{P}_j&=&\sum_{\alpha\in j}^{d+1}\hat{A}_{\alpha}-\hat{I}.
\end{eqnarray}
These imply
\begin{equation}\label{delta1}
 tr\{\hat{A}_{\alpha}\hat{P}_j\}=\Lambda_{\alpha,j};\;\;\;\Lambda_{\alpha,j}=\begin{cases}1,
  \;\;\alpha \in j, \\
                      0, \;\;\alpha \ni j.\end{cases}
\end{equation}

To prove Eq.(\ref{delta1}) we note Eq.(\ref{mub},  \ref{mub2}) to write,

\begin{eqnarray}
\hat{A}_{\alpha}\hat{P}_j&=&\hat{A}_{\alpha}+\sum_{\alpha'\ne
\alpha}\hat{A}_{\alpha'}\hat{A}_{\alpha}-\hat{A}_{\alpha}\;\;\;\alpha\in j \nonumber\\
&=&\sum_{\alpha'\ne\alpha}\hat{A}_{\alpha'}\hat{A}_{\alpha}-\hat{A}_{\alpha}\;\;\;
\alpha\ni j,
\end{eqnarray}
taking the trace implies (\ref{delta1}). This result, Eq.(\ref{delta1}), involves {\it
geometrical factors only}. We note that the $\Lambda_{\alpha,j}$ may equally be viewed as
\begin{equation} \label{delta2}
 tr\{\hat{A}_{\alpha}\hat{P}_j\}=\Lambda_{\alpha,j};\;\;\;\Lambda_{\alpha,j}=\begin{cases}1,
  \;\;j \in \alpha, \\
                      0, \;\;j \ni \alpha.\end{cases}.
\end{equation}
The proof for this case is given in Appendix A.,

e.g. for d=3 the underpinning's schematics is (the choice of c/2 for the CB vectors will prove convenient below).
\[ \left( \begin{array}{ccccc}
m\backslash b&-1&0&1&2 \\
0&A_{(c/2=0,-1)}&A_{(0,0)}&A_{(0,1)}&A_{(0,2)}\\
1&A_{(c/2=1,-1)}&A_{(1,0)}&A_{(1,1)}&A_{(1,2)}\\
2&A_{(c/2=2,-1)}&A_{(2,0)}&A_{(2,1)}&A_{(2,2)}\end{array} \right)\].\\

The geometrical line, $L_j, j=(1,2)$ given above (end of Section III) upon being transcribed
to its operator formula is via Eq.(\ref{mub2}),

\begin{equation}\label{lineop}
P_{j(c/2=1,m_0=2)}=A_{(c/2=1,-1)}+A_{(2,0)}+A_{(1,1)}+A_{(0,2)}-\hat{I}.
\end{equation}

Returning to Eqs.(\ref{ptop},\ref{mxel}), these equations imply that, the projection operators
$A_{\alpha}$,
in the CB representation are given by,

\begin{equation}\label{point}
\big(A_{\alpha=m,b}\big)_{n,n'}=\begin{cases}
\frac{\omega^s}{d};\;\;s=(n-n')(\frac{b}{2}[n+n'-1]-m),\;\;
b\ne-1,\\
\delta_{n,n'}\delta_{c/2,n}\;\;b=-1.\end{cases}
\end{equation}

First we argue that every $\big(A_{\alpha=m,b}\big)_{n,n'}$ in column b has in every other
column b' $b\ne b',\;b,b'\ne -1$ one projector such that
$\big(A_{\alpha=m,b}\big)_{n,n'}=\big(A_{\alpha'=m',b'}\big)_{n,n'}:$ Consider two distinct
columns, b,b' $(b,b'\ne-1)$ and given the matrix elements n,n' $(n\ne n')$ of a projector
$\big(A_{\alpha=m,b}\big)_{n,n'},$ compare it with $\big(A_{\alpha'=m',b'}\big)_{n,n'}.$ If
s (Eq.(\ref{point})) is $\ne s'$ i.e. $\frac{b}{2}(n+n')-1)-m \ne \frac{b'}{2}(n+n')-1)-m'$
pick another projector in the same column, b' (i.e vary m'). Since m' = 0,1,2...d-1 there
is one (and only one) $\big(A_{\alpha'=m',b'}\big)_{n,n'}$ such that
$\big(A_{\alpha=m,b}\big)_{n,n'}=\big(A_{\alpha'=m',b'}\big)_{n,n'}.$ Now consider another
matrix element $\big(A_{\alpha}\big)_{\bar{n},\bar{n}'}.$ We have trivially that
$\big(A_{\alpha}\big)_{\bar{n},\bar{n}'}=\big(A_{\alpha'}\big)_{\bar{n},\bar{n}'}$ iff
$\bar{n}+\bar{n}'=n+n'$. i.e. all matrix elements (n,n') with n+n'=c (constant) are such
that $\big(A_{\alpha}\big)_{n,n'}=\big(A_{\alpha'}\big)_{n,n'}.$ These elements are
situated along a line perpendicular to the diagonal of the matrices. We refer to this
perpendicular as FV (foliated vector), it is parameterized by  c.\\
We now assert that all other (non diagonal) matrix elements are unequal. i.e. for $b \ne
b'$, $\big(A_{\alpha=m,b} \big)_{n,n'}$ $\ne\;\big(A_{\alpha'=m',b'} \big)_{n,n'},\;\forall
\;n,n' \ni FV.$ Proof: Let two elements n,n' and l,l' with $n\ne n';\;l\ne l'$ in the two
matrices be equal. Thus (c=n+n', c'=l+l'):
\begin{eqnarray}
\frac{b}{2}(c-1)-m&=&\frac{b'}{2}(c-1)-m', \;\;\;and\nonumber \\
\frac{b}{2}(c'-1)-m&=&\frac{b'}{2}(c'-1)-m',\;\; \nonumber \\
\end{eqnarray}
These {\it imply} c=c', QED.  Now consider s=0. Then all the matrix elements along FV are
1/d. We have then that for $\big(A_{\alpha} \big)_{n,n'}=\big(A_{\alpha'}
\big)_{n,n'}=\omega^s/d,\;s \ne 0,$ d-1
matrix elements along FV are all distinct. The diagonal is common to all.\\
We have, thus, a prescription for d projectors, $A_{(m,b)},$ one for each b, ($b \ne -1),$
all having equal matrix elements along FV labelled by c. We supplement these with the
projector $A_{(c/2,-1)}=|c/2\rangle\langle c/2|$ to have the d+1 "points" constituting a
line j . $(|c/2 \rangle$ being a state in the CB.) Thus our line is formed as follows: It
emerges from $A_{(c/2,-1)}$ continues to $A_{(m(0),0)}$ in the b=0 column. Then it
continues to the points $A_{m(b),b}$ in succession: b=1,2...d-1 with m(b) determined by
$$\frac{b}{2}(c-1)-m(b)=\frac{b+1}{2}(c-1)-m(b+1).$$ Thus the two parameters, c=2m(-1) and m(1),
determine the line i.e. j=(m(-1),m(1). The general formula for the line is thus Eq.(\ref{m(b)}
now acquiring a meaning
in terms of the point operators, $A_{\alpha=m(b),b}$. It is of interest that, if we associate the
CB states with the position variable, q, of the continuous problem and its Fourier transform state,
viz b=0
 (cf. Eq.(\ref{mxel})), with the momentum, p, we have that the line of the finite dimension problem
 is parameterized with "initial" values of "q" and "p" i.e. m(-1) and m(0).\\
 The discussion of the properties of the line thus defined confirm that these lines realize DAPG
 lines. The analysis above indicate that the line operator, $P_{j=(m(-1),m(0))}.$ We now list some
 important consequences of this. We have shown that the matrix elements along a FV direction are
the same for all the point operators $A_{\alpha \in j}.$ Indeed that is how we defined our
lines. On the other hand we argued that the matrix elements {\it not} along the FV are all
distinct. Thence summing up d such terms residing on a fixed $P_j$ ({\it excluding the b=-1
and the diagonal term}) sums up for each matrix  element n,n' the d roots of unity for all
matrix elements not on FV, hence for all c,
\begin{equation}
\big(\sum_{\alpha \in j, \alpha \ni
\alpha_{-1}}^{d}\hat{A}_{\alpha}-\hat{I}\big)_{n,n'}=0;\;n,n'\ni
n+n'=c;\;\;\alpha_{-1}=|c/2\rangle\langle c/2|.
\end{equation}
Thus
 $\big(\hat{P}_j\big)_{n,n'}=\big(\sum_{\alpha\in j}^{d+1}\hat{A}_{\alpha}-\hat{I}\big)_{n,n'}\ne 0$
 {\it only} along FV, and is 1 along the diagonal at c/2=m(-1). The sum over the matrix
 elements on a FV, which are the same for all the $\hat{A}_{\alpha \in j, \ne -1}$ simply cancel
 the $1/d$. We illustrate this for
the example considered above Eq.(\ref{lineop}), viz: d=3, line with m(-1)=1, m(0)=2, i.e.
j=(1,2) Evaluating the point operators, $\hat{A}_{\alpha}$,
\begin{equation}\label{point1}
A_{(c/2=1,-1)}=\begin{pmatrix}0&0&0\\0&1&0\\0&0&0\end{pmatrix},A_{(2,0)}=\frac{1}{3}\begin{pmatrix}1&\omega^2&\omega\\
\omega&1&\omega^2\\\omega^2&\omega&1\end{pmatrix},A_{(1,1)}=\frac{1}{3}\begin{pmatrix}1&\omega&\omega\\\omega^2&1&1\\
\omega^2&1&1\end{pmatrix},A_{(0,2)}=\frac{1}{3}\begin{pmatrix}1&1&\omega\\1&1&\omega\\\omega^2&\omega^2&1\end{pmatrix},
\end{equation}
and evaluating the sum, Eq.(\ref{lineop}), gives
\begin{equation} \label{pj}
P_{j:(m(-1)=1,m(0)=2)}=\begin{pmatrix}0&0&\omega\\ 0&1&0\\ \omega^2&0&0\end{pmatrix}.
\end{equation}
Quite generally,
\begin{equation}\label{pc}
(P_{j=m(-1),m(0)})_{n,n'}=\begin{cases}\omega^{-(n-n')m(0)}\delta_{\{(n+n'),2m(-1)\}}\\
0\;\;otherwise. \end{cases}
\end{equation}
Thus,

\begin{equation}\label{p2}
(\hat{P}_{j=c/2,m_0}^2)_{n,n'}=\delta_{n,n'}.\;\;i.e.\;\hat{P}_j^2=\hat{I}\;\forall j .
\end{equation}

\begin{equation}\label{tm}
Theorem:\;\;
tr\hat{P}_j\hat{P}_{j'}=d\delta_{j,j'}.\nonumber
\end{equation}

\begin{eqnarray}
Proof:\;\;\;Eq.(\ref{p2})&\rightarrow&tr\hat{P}_j^2=tr \hat{I}=d, \nonumber \\
Eq(\ref{mub2})&\rightarrow&tr\hat{P}_j\hat{P}_j'=tr\big(\hat{P}_j\sum_{\alpha\in j'}\hat{A}_{\alpha}\big)-tr\hat{P}_j \nonumber \\
Eq(\ref{delta1})&\rightarrow& tr\hat{P}_j\hat{A}_{\beta\in j}-1=0,\;\;j'\ne j.\nonumber
\end{eqnarray}
Where we used $tr\hat{P}_j=1$ as follows from Eq.(\ref{p2}) and that two distinct lines share one
 point (see DAPG(b)).

\section{   Further Attributes of the line operator, $\hat{P}_j$}

The line operator, $\hat{P}_j$ is hermitian being the sum of hermitian operators. Noting that, trivially, $tr\hat{P}_j=1$ and $\hat{P}^2=\hat{I},$  Eq(\ref{p2}) implies the spectral representation
 \begin{equation}
 (\hat{P_j})=\begin{cases}\delta_{n,n'}\;\;n=1,2,...\frac{d+1}{2}\;\;(\frac{d+1}{2}\;elements),\\
-\delta_{n,n'}\;\;n=\frac{d+3}{2},\frac{d+5}{2},...,d\;\;(\frac{d-1}{2}\;elements.)
\end{cases}
\end{equation}
We thus have that
\begin{equation}\label{p3}
\mathbb{P}_j\equiv \frac{\hat{P_j}+\hat{I}}{2}
\end{equation}
is a projection operator onto the $\frac{d+1}{2}$ eigenstates of $\hat{P_j}$ with
eigenvalue +1. Recalling that $\hat{P_j}=\sum_{\alpha \in j}\hat{A_\alpha}-\hat{I}$ and
$\hat{P}^2=\hat{I}$ implies (what we term) the Fluctuation Distillation Formula (FDF):
\begin{equation}\label{fdt}
\sum_{\alpha \ne \beta;\alpha,\beta \in j} {\hat{A}_{\alpha}\hat{A}_{\beta}}=\sum_{\alpha
\in j}{\hat{A}_{\alpha}}.
\end{equation}
The proof is given in Appendix B.\\

 Summing both sides of the Eq.(\ref{delta1}) over
$\alpha$ and use Eq.(\ref{mub2}) and Eq.({\ref{mub}) to write, for the LHS
\begin{eqnarray}\label{lhs}
tr\big[\big(\sum_{\alpha\in j}^{d+1}\hat{A}_{\alpha}\big)\big(\sum_{\alpha' \in
j}^{d+1}\hat{A}_{\alpha'}-\hat{I}\big)\big]&=&
tr\big[\sum_{\alpha\ne\beta}\hat{A}_{\alpha}\hat{A}_{\beta}\big] \nonumber \\
                    &=&tr\big[(\hat{P}_j + \hat{I})\hat{P}_j \big] = tr[\hat{P}_j + \hat{I}].
\end{eqnarray}
Where in the last step we used the relation, Eq.(\ref{p2}),  $\hat{P}_j^2=\hat{I}$ (valid for
$A_{\alpha_{-1}}=|c/2\rangle\langle c/1|$).

To illustrate the spectral decomposition we consider again the line $j=\{m(-1)=1),m(0)=2\}$ in
d=3. This line runs through the  points (1,-1),(2,0),(1,1) and (0,2). The line operator $\hat{P}_(1,2)$ was given above,
\begin{equation}
P_{1,2)}=\begin{pmatrix}0&0&\omega\\ 0&1&0\\ \omega^2&0&0\end{pmatrix}=
\begin{pmatrix}1/2&0&\omega/2\\0&1&0\\ \omega^2/2&0&1/2\end{pmatrix}-\begin{pmatrix}1/2&0&\omega/2\\0&1&0\\-\omega^2/2&0&1/2\end{pmatrix}.
\end{equation}
Thus the matrix is diagonal in the orthonormal basis,
$$
\left(\begin{array}{c}\omega/{\sqrt2}\\0\\\omega^2/{\sqrt2}\end{array}\right);\;
\left(\begin{array}{c}0\\1\\0\end{array}\right);\;
\left(\begin{array}{c}\omega/{\sqrt2}\\0\\-\omega^2/{\sqrt2}\end{array}\right).
$$
The associated projection operator,
\begin{equation}
\mathbb{P}_{(1,2)}= \frac{\hat{P_j}+\hat{I}}{2}=\frac{1}{\sqrt2}[\omega^2|0\rangle+\omega|2\rangle]
\frac{1}{\sqrt2}[\omega\langle0|+\omega^2\langle2|]+|1\rangle\langle1|.
\end{equation}

Let $\Bbb{M}_j$ be a d(d+1) matrix. Its d+1 columns are made of the d elements, $\langle n|m(b),b\rangle$ with $|m(b),b\rangle$ the state whose projection, $|m(b),b\rangle\langle b,m(b)|,$
is the point of the line j in the column b. Define $\Bbb{M}_j^{\dagger}$ as the d
 columns matrix whose rows are made up of d+1 columns corresponding to the adjoint of
  $\Bbb{M}_j$. For example, using Eq.(\ref{lineop}) the line $j:c/2=1,m(0)=2$ is given by the
 aggregate of points: $(c/2=1,m(0)=2),(2,0),(1,1),(0,2)$ giving as a normalized state line operator,

 $$\Bbb{M}_j=\frac{1}{\sqrt{d+1}}\left ( \begin{array}{cccc}
 0&1/\sqrt3&1/\sqrt3&1/\sqrt3\\
 1&\omega/\sqrt3&\omega^2/\sqrt3&1/\sqrt3\\
 0&\omega^2/\sqrt3&\omega^2/\sqrt3&\omega^2/\sqrt3\end{array} \right),
\Bbb{M}_j^{\dagger}=\frac{1}{\sqrt{d+1}}\left ( \begin{array}{ccc}
    0&1&0\\
1/\sqrt3&\omega^2/\sqrt3&\omega/\sqrt3\\
1/\sqrt3&\omega/\sqrt3&\omega/\sqrt3\\
1/\sqrt3&1/\sqrt3&\omega/\sqrt3\end{array} \right).$$
While the line $j=c/2=0,m_0=0:$ viz
$$(c/2=0,m(0)=0),(0,0),(1,1),(2,2)\Rightarrow \Bbb{M}_{(c/2=0,m(0)=0)}=\frac{1}{\sqrt 4}\left(
\begin{array}{cccc}
1&1/\sqrt3&1/\sqrt3&1/\sqrt3\\
0&1/\sqrt3&\omega^2/\sqrt3&\omega/\sqrt3\\
0&1/\sqrt3&\omega^2/\sqrt3&\omega/\sqrt3\end{array} \right)$$

 The construction of the matrices assures
\begin{eqnarray}
tr {\Bbb{M}_j^{\dagger}\Bbb{M}_j}&=&1\;\;\nonumber \\
tr {\Bbb{M}_j^{\dagger}\Bbb{M}_{j'}}&=&\frac{1}{d+1}\;\;j\ne j'.
\end{eqnarray}
The proof is based on the DAPG attribute that two distinct lines share precisely one point,
and for equal b distinct states are orthogonal. e.g.
$$tr\Bbb{M}_{(c/2=0,m(0)=0)}^{\dagger}\Bbb{M}_{(c/2=1,m(0)=2)}=\frac{1}{4}.$$\\
The normalized "line" matrix, $\Bbb{M}_j$, is the "square root" of the normalized "line"
projection operator, viz:
\begin{equation}
\Bbb{M}_j\Bbb{M}^{\dagger}_j=\frac{2}{d+1}\Bbb{P}_j.
\end{equation}
This is almost self evident: The RHS equals
$$\sum_{\alpha \in j}A_{\alpha}=\sum_{b}|m(b),b\rangle\langle b,m(b)|$$
While the the LHS is (for the n,n' matrix element)
$$\sum_{b}\langle n|m(b),b\rangle\langle b,m(b)|n'\rangle.$$
i.e. they are identical.\\
We now demonstrate that these line operators are geometric in origin. Thus  we associate the
line operator  $\Bbb{M}_j$ with  $\tilde{\Bbb{M}}_j,$ defined by  the replacement, in the former, of every column b  ($\ne -1$), as follows. Instead of the elements $\langle n|m(b),b\rangle,\;n=0,1,...d-1$ with m(b) a point  in the line j, we have in the  column b of $\tilde{\Bbb{M}}_j,$  1 at the row corresponding to m(b). Thus in our example of j=(1,2) for d=3 (cf. Eq.(\ref{lineop})), we have,

$$\Bbb{M}_{(1,2)}=\frac{1}{\sqrt{4}}\left ( \begin{array}{cccc}
 0&1/\sqrt3&1/\sqrt3&1/\sqrt3\\
 1&\omega/\sqrt3&\omega^2/\sqrt3&1/\sqrt3\\
 0&\omega^2/\sqrt3&\omega^2/\sqrt3&\omega^2/\sqrt3\end{array} \right)
\Rightarrow\tilde{\Bbb{M}}_{(1,2)}=\frac{1}{\sqrt{4}}\left ( \begin{array}{cccc}
 0&0&0&1\\
 1&0&1&0\\
 0&1&0&0\end{array} \right).$$
$\tilde{\Bbb{M}}_j$ is geometrical: it involves, in essence, the drawn line. We note that
the transition between $\Bbb{M}_j$ and $\tilde{\Bbb{M}}_j$ is "local" unitary
transformation: we require a distinct unitary transformation for each column (basis),b.

\section{   DAPG Underpinned Quasi-Distributions}

We now define a Wigner function like quasi-distribution, $V_{\rho}[j=(m(-1),m(0))].$ It maps the density operator $\hat{\rho}$ onto the lines, $j=\{(m(-1),m(0))\}$, and points $\alpha=\{m(b)\}$, of DAPG. This, $V_{\rho}(j)$, Wigner function-like completely determine and is determined by the state of the system,$\hat{\rho},$. ( An essentially equivalent definition relates to arbitrary operators in the d- dimensional Hilbert space.) We then consider summation of  $V_{\rho}(j)$ along    a fixed values of $\alpha$ that represent an MUB projector, as the finite dimensional Radon transform
of  $V_{\rho}(j)$. The inversion of this, viz. the reconstruction of $V_{\rho}(j)$ from summations along such points is our finite dimensional inversion of the Radon transform.\\

Returning to the line operators, $\hat{P}_j,\;\;j=1,2,...d^2,\;\;j\equiv(m(-1),m(0))$,
Eq.(\ref{m(b)}), we utilize their orthogonality, $tr\hat{P}_j\hat{P}_{j'}=d\delta_{j,j'},$
to write,

\begin{equation}\label{vj}
\hat{\rho}\;=\;\ \frac{1}{d}\sum^{d^2}_j\big(tr \hat{\rho}\hat {P}_j \big)\hat{P}_j=
\frac{1}{d}\sum^{d^2}_jV_{\rho}(j)P_j.
\end{equation}
Where we defined $V_{\rho}(j)\equiv tr\big(\hat{\rho}\hat{P}_j \big)$. A partial list of
attributes
of $V_{\rho}(j)$ is the following.\\
\noindent$1.\;V_{\rho}(j)=\big(V_{\rho}(j)\big)^{\ast}.$\\
\noindent$2.\;\frac {1}{d}\sum^{d^2}_j V_{\rho}(j)=tr\rho \big(\frac{1}{d}\sum^{d^2}_j
\hat{P}_j\big)=1,\;\;\;cf Eq.(\ref{mub2}).$ \\
\noindent$3.\;tr \hat{\rho}= \frac{1}{d}\sum^{d^2}_j\big(tr\hat{\rho}\hat{P}_j
\big)tr\hat{P}_j=
\frac{1}{d}\sum^{d^2}_j\big(tr \hat{\rho}\hat{P}_j \big)=1.$\\
\noindent$4.\;tr\hat{A}\hat{B}=\frac{1}{d}\sum^{d^2}_j\big(tr\hat{A}\hat{B}\hat{P}_j\big)tr\hat{P}_j=
\frac{1}{d}\sum^{d^2}_jV_{AB}(j).$\\
\noindent$5.\;tr\hat{A}\hat{B}=\frac{1}{d^2}\sum^{d^2}_j\big(tr\hat{A}\hat{P}_j\big)\sum^{d^2}_{j'}
\big(tr\hat{B}
\hat{P}_{j'}\big)tr\hat{P}_j\hat{P}_{j'}=\frac{1}{d}\sum^{d^2}_jV_{A}(j)V_{B}(j).$\\

In finite dimensional studies it is convenient to use unitary operators
\cite{weyl,schwinger}. Thus the probability, given that the system is in the state $\hat{\rho},$ to measure it to be in
the state $|m,b\rangle$ i.e. to be in an eigen function of  $XZ^b$ with eigenvalue $\omega^m$ without regard to any other probability is
$tr\big(\hat{\rho} |m,b\rangle \langle b,m|\big)$. Here $|m,b\rangle$ is the eigen-function of
the unitary operator $XZ^b$, cf Eq.(\ref{tal1}).
 $tr\big(\rho |m,b\rangle \langle b,m|\big)$ is the finite dimensional observable that
 corresponds to $tr\big(\hat{\rho} |x',\theta\rangle \langle \theta,x'|\big)$ of Eq. (\ref{r}) in the
 continuous case. Note that we may regard the c-number function gotten upon mapping, a la Wigner, the operator $\hat{\rho}|x',\theta\rangle \langle \theta,x'|$ onto phase space as a marginal quasi distribution of $W_{\rho}(q,p)$  cf. Eq.(\ref{g}). In what follows we introduce, for the finite dimensional case, Wigner like mapping, now onto DAPG coordinates, that, correspondingly, relates its marginals to the full quasi distribution for the MUB state projectors.\\
 Recalling, Eq.(\ref{ptop}), $\hat{A}_{\alpha}\equiv|m,b\rangle \langle m,b|,\;\;\alpha=(m,b)$,
 we may write,
\begin{equation}\label{rad2}
tr\hat{\rho}\hat{A}_{\alpha}=\frac{1}{d}\sum^{d^2}_jV_{\rho}(j)V_{A_{\alpha}}(j),\;\;cf.\;
4 \;above.
\end{equation}
Noting that,
\begin{equation}
V_{A_{\alpha}}(j)=tr A_{\alpha}P_j=\Lambda_{\alpha,j}\;\;cf. \;Eq.(\ref{delta1}).
\end{equation}
i.e.
\begin{equation}
tr\hat{\rho}\hat{A}_{\alpha}=\frac{1}{d}\sum_{j\in
\alpha}tr\hat{\rho}\hat{P}_j=\frac{1}{d}\sum_{j}tr\hat{\rho}\hat{P}_j\Lambda_{\alpha,j}=
\frac{1}{d}\sum_{j}V_{\rho}(j)V_{A_{\alpha}}(j).
\end{equation}
These equations correspond to Eqs.(\ref{g}),(\ref{r}) of Section II. Thus we identify $tr\rho
A_{\alpha}$ as the (finite dimensional) Radon transform of $V_{\rho}(j)$ - it sums up the
values of $V_{\rho}(j)$ for $j\in\alpha$. (The $\Lambda_{\alpha,j}$ plays the role of the delta function.) These equations correspond to Eq.(\ref{g},  \ref{w}). \\
To obtain the (finite dimensional) inversion to the transform we consider,
\begin{equation}
\sum_{\alpha \in j'}tr\big(\rho \hat{A}_{\alpha}\big)=\frac{1}{d}\sum_j tr \rho P_j
\sum_{\alpha \in j'}\Lambda_{\alpha,j}=\frac{1}{d}\sum_{j\in\alpha}\sum_{\alpha \in j'}tr
\rho P_j=1+V_{\rho}(j).
\end{equation}
Where in the last step we used the DAPG based relation,
$$\sum_{(\alpha \in j)}\sum_{(j' \in
\alpha)}\hat{P}_j=\sum_{j'=1}^{d^2}\hat{P}_{j'}+dP_j=d\hat{I}+d\hat{P}_j.$$ Thence, the
inversion is
\begin{equation}
V_{\rho}(j)=\sum_{\alpha \in j}tr\big(\rho \hat{A}_{\alpha}\big)-1.
\end{equation}
This could have been gotten directly from the definition of the operators however we could
perhaps miss thereby some of the insight that the lengthy derivation provides which, in turn,
underscores
its relation with the continuous inverse Radon transform, Eq.(\ref{y}). \\

\section{  Affine Plane Geometry (APG) Formulation}

The central work in the underpinning of finite dimensional MUB operators with finite
geometry, \cite{wootters4}, is given in terms of lines and points of affine plane geometry
(APG) rather than our choice of DAPG. An advantage of this, APG,  scheme is its apparent
similarity with the $d\rightarrow \infty$ case in that it involves square arrays and states
projectors are straight lines (albeit in a modular sense) in a "discrete" phase space. It
allowed a direct imposition of translational invariance  and extension to dimensionality
d=$p^n;\;n>1,\;p\;a \;prime$, \cite{wootters3}. In this section we recast our DAPG
underpinning into an APG one by interchanging lines and points and in particular the
symmetrical meaning of $\Lambda_{\alpha,j}$, Eq. (\ref{delta1}),(\ref{delta2}) is shown to
allow the {\it formulae} for the Radon transform remain intact.\\
Recall that within the DAPG underpinning a line was defined by the two points: m(-1) and
m(0). The first, m(-1), was associated with modulated position, as it relates to the eigen
values of Z, viz the computational basis states. We shall refer to it by $\xi$. The second,
m(0), was associated with (modulated) momenta - it being the eigen values of X (b=0,
Eq.(\ref{tal1}), i.e. the Fourier transform of the former states. We shall refer to it by
$\eta$. Thus a DAPG lines are $j=(m(-1)\equiv \xi,m(0)\equiv
\eta);\;\xi,\eta=0,1,2,...d-1.$ Now consider a $d \cdot d$ square array, $d=p,\;a prime
\ne2$, whose (discrete) coordinates along the x axis is labelled by $\xi$ and the y axis by
$\eta$. We interpret each point, $\alpha$, in the array, $\alpha=(\xi, \eta)$ as
underpinning a DAPG line j. Thus the "image" of each DAPG line is a APG point. We now
consider lines in this array:
 \begin{eqnarray}\label{apgl}
 \eta &=& r\xi + s;\;mod[d];\;\;r,s=0,1,2,...d-1. \nonumber \\
 \xi  &=& s';\;\;mod[d];\;\;s'=0,1,2,...d-1.
 \end{eqnarray}
 Eq.(\ref{apgl}) defines d(d+1) "straight" lines: there are $d^2$ possibilities for r and s
 and d for s'. Each line contain d points:  $(\xi_i,\eta_i),\;i=0,1,...d-1.$ The line $xi=s'$
contain d points as well: $ (s',\eta_i),\;i=0,1,...d-1.$ this proves APG(a), of section
III. Similarly, the proofs of the validity of APG(x), x=b,c,d and e for the lines,
Eq.(\ref{apg1}) and points forming the array are trivial. e.g. to prove APG(b): Consider
two distinct lines, $\eta=r_1\xi+ s_1,\;\eta=r_2\xi+s_2,\;r_1\ne r_2,\;s_1 \ne s_2.$ Let
this lines share a point $(\xi_0,\eta_0)$. This implies, $r_1\xi_0+s_1=r_2\xi_0+s_2.$  This
implies a {\it unique} $(\xi_0,\eta_0):\;\eta_0=r_1\xi_0+s_1;\;\xi_0=(s_1-s_2)/(r_2-r_1).$
For $r_1=r_2,\;s_1\ne s_2$, no common point exist (the lines are parallel). Thus the square
array with points labelled by $(\xi,\eta)$ and lines given by Eq(\ref{apgl}) form a
realization of APG. It is specified that a line $j=(m(-1),m(0))\in DAPG$ is mapped to a
point $(\xi,\eta) \in APG.$ (Note that a point, $\alpha$,
of DAPG underpins an MUB projector: $\alpha=|m,b\rangle \langle b,m|$.)\\
Theorem: In a DAPG realization, the d lines of DAPG that form the image of the d points of a APG line,
$\eta=r\xi+s$, share a point.\\
Proof: Consider an APG line, $\eta=r\xi+s$. It contain the d APG points
$(\xi_i,\eta_i=r\xi_i+s),\;i=0,1,..d-1.$ Now pick two arbitrary points i,i'. Their images
in DAPG are the two lines $(m(-1)=\xi_i,m(0)=r\xi_i+s)$ and
$(m(-1)=\xi_{i'},m(0)=r\xi_{i'}+s).$ The equation for the point they share, cf
Eq.(\ref{twopoints}), is
\begin{eqnarray}\label{trans}
\frac{b}{2}(2\xi_i-1)+r\xi_i+s&=&\frac{b}{2}(2\xi_{i'}-1)+r\xi_{i'}+s\;\;mod\;[d], \nonumber \\
\rightarrow\;&\;&(b+r)(\xi_i-\xi_{i'})=0\;\;mod\;[d].
\end{eqnarray}
this is independent of $\xi_i-\xi_{i'}$. i.e. all the lines, i,i'=1,2,..d-1, share a common point at
b=-r mod [d], thence the point is m(b=-r)=r/2+s. For the APG line $\xi=s'$ the common point within DAPG is,
trivially, at b=-1: m(-1)=s'.\\
To illustrate the steps involved we consider d=3 with APG line given by $\eta=\xi+1$. Thus the APG points
involved are: (0,1),(1,2) and (2,0). Via Eq(\ref{pj}), Eq.(\ref{pc}) we have,
\begin{equation}
P_{(0,1)}=\begin{pmatrix}1&0&0\\ 0&0&\omega\\ 0&\omega^2&0\end{pmatrix};\;\;
   P_{(1,2)}=\begin{pmatrix}0&0&\omega\\ 0&1&0\\ \omega^2&0&0\end{pmatrix};\;\;
P_{(2,0)}=\begin{pmatrix}0&0&\omega\\ 0&1&0\\ \omega^2&0&0\end{pmatrix}.
\end{equation}
Now, Eq.(\ref{trans}) relates these to the DAPG point (i.e. the MUB projector)
$|0,2\rangle \langle2,0|=A_{(0,2)}$
\begin{equation}
\frac{1}{3}\Big[\begin{pmatrix}1&0&0\\ 0&0&\omega\\ 0&\omega^2&0\end{pmatrix}+\begin{pmatrix}0&0&\omega\\ 0&1&0\\ \omega^2&0&0\end{pmatrix}+ \begin{pmatrix}0&0&\omega\\ 0&1&0\\ \omega^2&0&0\end{pmatrix}\Big]=\frac{1}{3}\begin{pmatrix}1&\omega^2&\omega\\ \omega&1&\omega^2\\ \omega^2&\omega&1\end{pmatrix},
\end{equation}
where the last matrix is $A_{(0,2)},$ cf. Eq.(\ref{point1}).\\

To distinguish between DAPG and APG underpinned operators we adopt the following scheme. The projector
onto the MUB state $|m,b\rangle$, a point $al\grave{a}$ DAPG, was designated by $A_{(\alpha=m(b))}$
with the subscript indicating its coordinates. It is a line $al\grave{a}$ APG, and will be designated
by $B_{(\lambda=\eta(\xi))}$ the subscript now gives the APG line's equation. The DAPG line operator
$P_j$  will, within APG underpinning scheme, be specified by the point $P_{(\xi,\eta)}$ giving its
coordinates. Within this notation our Wigner function-like mapping function, Eq.(\ref{vj}), is,
\begin{equation}\label{w3}
\cal{V}_{\rho}(\xi,\eta)=tr \rho Q_{(\xi,\eta)}.
\end{equation}
Here the variables, $\xi, \eta$, signifies its APG underpinnings. In conformity with APG we have,
(in correspondence with the DAPG relation, Eq.())
\begin{equation}\label{bq}
B_{(\eta=\eta(\xi))}=\frac{1}{d}\sum_{(\eta,\xi \in \beta)}Q_{(\eta,\xi)}.
\end{equation}
For the sake of clarity we wish to iterate the relations between the {\it Hilbert space}
operators $P_j$ and $Q_{\xi,\eta}$: $P_j$ is a line operator within DAPG underpinning,
j=(m(-1),m(0)) - the line is fully parameterized by the "position" variable value, m(-1)
and the "momentum" m(0). $Q_{\xi,\eta}$ is a APG underpinned point operator with $\xi,\eta$
the point's coordinates with $\xi=m(-1),\eta=m(0)$. The APG underpinning was {\it
constructed} by identifying,
$$P_{j=(m(-1),m(0))}\equiv Q_{(\xi=m(-1),\eta=m(0))}.$$
Thus although the subscript of P designate a {\it line} underpinning (DAPG); while that of Q, is a
{\it point} underpinning (APG), they are equal when the subscripts are numerically equal.\\

The essential issue in the following is the simple, yet long winded, equivalence of our two accounts for
the MUB operator, $|m,b\rangle \langle b,m|$:\\
\noindent(a) As a DAPG underpinned point, $\alpha$. In this case $\alpha=m(b)$, specifies
the coordinate of the point in the $d\cdot(d+1)$ points array. It is given by the point
underpinned operator $A_{\alpha}$.\\
\noindent(b) As a APG underpinned line, $\lambda$. In this case $\lambda=\eta(\xi)$ is the
equation of the line whose constituent points, $\xi,\eta \in \lambda$, correspond to the
DAPG underpinned lines parameterized with m(-1), "position", and m(0), "momentum"
coordinate: $m(-1)\rightleftarrows\xi;\;m(0)\rightleftarrows\eta$. It is given by the line
underpinned operator
$B_{\lambda}$.\\
The mapping of the line operator, $B_{\lambda},$ is the Lambda function,
\begin{equation} \label{delta3}
tr B_{\lambda}Q_{\xi,\eta}=\frac{1}{d}\sum_{\xi',\eta' \in \lambda}^d
Q_{\xi',\eta'}Q_{\xi,\eta}= \Lambda_{((\xi',\eta'), \lambda)}=\begin{cases}1,
  \;\;(\xi',\eta') \in \lambda, \\
                      0, \;\;(\xi',\eta') \ni \lambda.\end{cases}. QED.
\end{equation}
i.e. it is non-vanishing only with  the point $(\xi,\eta)$ in the line $\lambda$. It is a straight
forward matter to show that the mapping of the density operator onto APG phase space like lines and
points, $\cal{V}_{\rho}(\xi,\eta)$, is a quasi distribution - i.e. it possess the equivalent attributes,
 1 - 5, heeded by our DAPG mapping given in the last section. AS an example  we prove item 5:
\begin{eqnarray}
&=&\frac{1}{d^2}\sum_{\xi,eta}^{d^2}\big[tr AQ_{\xi,\eta}\big]\sum_{\xi',\eta'}^{d^2} \big[tr BQ_{\xi',\eta'}\big]tr Q_{\xi,\eta}Q_{\xi',\eta'} \nonumber \\
     &=&\frac{1}{d}\sum_{\xi,\eta}\big(tr AQ_{\xi,\eta}\big)\big(tr BQ_{\xi,\eta}\big)=\frac{1}{d}\sum_{\xi,\eta}
\cal{V}_{A}(\xi,\eta)\cal{V}_{B}(\xi,\eta).
\end{eqnarray}
The Radon transform of the quasi distribution, $V_{\rho}(\xi,\eta),$ is $tr \rho B_{\lambda}$ - since in the present case the line underpinned operator is the MUB state projector. Thus,
\begin{equation}\label{exp}
tr\rho B_{\lambda}=\frac{1}{d}\sum_{\xi,\eta \in \lambda}^d tr\rho Q_{\xi,\eta}=\frac{1}{d}\sum_{\xi,\eta}\big(tr \rho Q_{\xi,\eta}\big)\Lambda_{(\xi,\eta),\lambda}.
\end{equation}
This expression is in complete analogy with Radon transform in the continuum: we sum the
quasi distribution over points $(\xi,\eta)$ on the line $\lambda$. Given that the system is
the state $\rho$, the probability to measure it to be in $|m,b\rangle$, is given by
Eq.(\ref{exp}). This equation is analogous to Eq.(\ref{g}): It expresses the probability of
being in the MUB state (here $|m,b\rangle$) in terms of summation of the quasi distribution
(here $tr \rho Q_{\xi,\eta}$) along a line determined by the mapping of the MUB projector
onto the underpinning points (here $\Lambda_{(\xi,\eta),\lambda}= tr
B_{\lambda}Q_{\xi,\eta}$).

\section {   Summary and Concluding Remarks}

The finite dimensional,d, density operator, and mutual unbiased basis (MUB) states' projectors were
mapped onto points and lines of finite geometry. These mappings were shown to be analogous to the
Wigner function mapping of the density operator and MUB state projectors in the continuum,
$d\rightarrow \infty$. In the latter ($d\rightarrow \infty$) case, the expectation values of the MUB
state projectors expressed via the appropriate Wigner function scheme were observed to be the Radon
transform of the Wigner function of the state itself. Inverse Radon transforms is, thus,  state
reconstruction in terms of the afore mentioned expectation values.\\
The proposed finite dimensional map of the density operator possess the quasi distributional attributes
over its underpinning geometry as the Wigner function over phase space. The mapping of the MUB state
projectors are, like their Wigner function counters, lines. These lines are arranged as straight lines
for the affine plane geometry (APG) underpinning. The underpinning with the dual affine plane  geometry
(DAPG) which was most extensively employed allows a simpler formulation.\\
The expectation values of the MUB state projectors were used to define finite dimensional Radon transform.
It involve summation along a line of the underpinning geometry. The inverse finite dimensional Radon
transform is used for state reconstruction in close analogy with the continuum
analysis.

A brief summary of dual affine plane geometry (DAPG) in finite dimension, d, was given. The
geometry was used to underpin projectors of states of mutual unbiased bases (MUB) scanning a
d-dimensional Hilbert space. The dimensionality studied were d=p, a prime, $\ne 2$.\\
The Wigner function, $W_{\rho}(q,p)$, may be viewed as a mapping of the density operator that act in
Hilbert space onto a c-number function in phase space. A finite dimensional, d, Wigner function-like,
 $V_{\rho}(j)$,  was defined a a mapping of the finite dimensional density operator, $\rho$, onto
 c-number function of lines, $L_j=1,2,...d^2$ and points, $\alpha=1,2...d(d+1)$, of dual affine
 plane geometry in d- dimensions. $V_{\rho}(j)$ posses all the attributes of $W_{\rho}(q,p)$ that
 qualify
it as quasi distribution. A particularly attractive attribute of the Wigner function that underscores
its role as quasi distribution is it marginals. In particular the marginals of mutual unbiased basis
 (MUB) state projector is itself a quasi distribution and is recognized as a Radon transform of
 $W_{\rho}(q,p)$. This transform involves integration along a line in phase space which relates to
 the MUB state projector map onto this space. In close analogy the marginal of a finite dimensional
 MUB state projector involves summation along a DAPG line ans is itself a quasi distribution.
 This close analogy led, through its physical interpretation, to circumvention the
angular involvement within Radon transformation and allowed a convenient definition of finite
dimensional Radon transform concomitant with its inverse i.e. the state reconstruction.\\
In closing we present in a comparative way the mappings involved in this work. (The
detailed meaning of the symbols are given in the text with the specified equations.)\\
\noindent (a)Continuum Hilbert space operator $\hat{A}$ onto phase space (Wigner function).
Here a "line" underpins an Mutual Unbiased Basis (MUB) state projector
$|x',\theta\rangle\langle\theta.x'|$. Eq.(\ref{w1}).\\
\noindent (b) Finite dimension, d=prime$\;\ne2$ Hilbert space operator $\hat{A}_N$ onto
points and lines of Dual Affine Plane Geometry (DAPG). Here a "point" $\alpha=m(b)$,
underpins an MUB state projector $|m,b\rangle\langle b,m|$.\\
\noindent (c) Finite dimension, d=prime$\;\ne2$ Hilbert space operator $\hat{A}_N$ onto
points and lines of Affine Plane Geometry (APG). Here a "line", $\lambda=(m(-1),m(0))$,
underpins MUB state projector $|m,b\rangle \langle b,m|$:\\

\noindent (a)$W_A(q,p)=\int dy e^{ipy}\langle q-y/2|\hat{A}|q+y/2\rangle.$  Eq.(\ref{w1})\\

\noindent (b)$V_{A_N}(j)= tr\hat{A}_N P_j;$  j - a line. Eq.(\ref{vj}).\\

\noindent (c)${\cal{V}_{A_N}}(\xi,\eta) = tr\big(\hat{A}_N Q_{\xi,\eta}\big);\;\;\;(\xi,\eta)$ - a point. Eq.(\ref{w3}).\\

Likewise we give now the Radon transform of the density operator mappings.\\

\noindent (a)$tr \big(\rho |x',\theta\rangle
\langle\theta,x'|\big)=({\cal{R}}W_{\rho})(x',\theta)=\\$
  =$\int \frac{dqdp}{2\pi}\delta(x'-qcos\theta-psin\theta)W_{\rho}(q,p).
\;\delta(x'-qcos\theta-psin\theta)=W_{|x'\theta \rangle\theta,x'|}(q,p).\;\; Eq.(\ref{r}).$\\

 \noindent (b)$tr\rho\hat{A}_{\alpha}=({\cal{R}_N}V_{\rho})(\alpha=m(b))=\frac{1}{d}\sum_j^{d^2}\big(tr\rho
 P_j\big)\Lambda_{\alpha,j};\;\Lambda_{\alpha,j}=V_{A_{\alpha}}(j).\;\;\;Eq.(\ref{rad2}).$\\

 \noindent(c)$tr\rho\hat{B}_{\lambda}=({\cal{R}_N}{\cal{V}_{\rho}})(\lambda=\eta(\xi))=\frac{1}{d}\sum_{\xi,\eta}\big(tr\rho Q_{\xi,\eta}\big)
 \Lambda_{(\xi,\eta),\lambda};\;\Lambda_{(\xi,\eta),\lambda}={\cal{V}_{B_{\lambda}}}(\xi,\eta).\;\;\;
Eq.(\ref{exp}).$

\section*[Appendix A] {Appendix A: The lambda function, $\Lambda_{(\alpha,j)}$}

To prove the alternative meaning of the Lambda function, Eq.(\ref{delta2}), consider
\begin{equation}
tr\hat{A}_{\alpha}\hat{P}_j=\frac{1}{d}\sum_{j'\in
\alpha}tr(P_jP_{j'})=\Lambda_{j,\alpha}.\nonumber
\end{equation}
Here we used the expression for $A_{\alpha}$ in terms of $P_j$, Eq.(\ref{oprel3}).
Evaluating this gives,

\begin{equation} \label{delta2}
 tr\{\hat{A}_{\alpha}\hat{P}_j\}=\Lambda_{\alpha,j};\;\;\;\Lambda_{\alpha,j}=\begin{cases}1,
  \;\;j \in \alpha, \\
                      0, \;\;j \ni \alpha.\end{cases}. QED.
\end{equation}
Where we used Eqs.(\ref{p2},  \ref{tm}).

\section*[Appendix B]{Appendix B: Fluctuation Distillation Formula}

Given,  Eq(\ref{oprel3}),   $\hat{P_j}=\sum_{\alpha \in j}\hat{A_\alpha}-\hat{I}$ and,
  Eq(\ref{p2}), $\hat{P_j}^2=\hat{I},$ implies

$$\big( \sum_{\alpha \in j}\hat{A_\alpha}-\hat{I}\big) \big(\sum_{\alpha' \in j}\hat{A_\alpha'}
-\hat{I} \big)=\hat{I}.$$ Thus,
$$\sum_{\alpha,\alpha' \in j}\hat{A_\alpha}\hat{A_\alpha'}=2\sum_{\alpha \in j}\hat{A_\alpha}.$$
Recalling that, Eq(\ref{ptop}), $A_{\alpha}^2=A_{\alpha}$ allows
$$\sum_{\alpha \ne\alpha' \in
j}\hat{A_\alpha}\hat{A_\alpha'}=\sum_{\alpha \in j}\hat{A_\alpha}.$$ QED

\section*[Appendix C] {Appendix C: The relation $\hat{P}_{j=m(-1),m(0)}^2=\hat{I}$}

Recalling Eqns.(\ref{lineop}, \ref{point}, \ref{pc}) we have
\begin{equation}
(\hat{P}_{j=c/2,m(0)})_{n,n'}= \begin{cases}\omega^{-(n-n')m(0)}\delta_{(n+n'),c}\\
0\;\;otherwise\end{cases} .
\end{equation}
Squaring this matrix we have trivially 1 at n=n', n=c/2. The only non nil element of the
n-th row ($n \ne c/2$) of the matrix is along the column n'=c-n, where it is given by
$\omega^{(n'-n)m(0)}$. The only column having non nil element at the row n=c-n' is the n-th
column with the element $\omega^{(n-n')m(0)}$. Thus
$(\hat{P}_{j=c/2,m(0)}^2)_{n,n'}=\delta_{n,n'}$ QED.\\

 Acknowledgments: I have greatly benefited from Prof T. Etzioni's guidance through the
 labyrinth of finite geometry mathematics and from discussions with Prof. A. Mann. In the early stages of this work I benefited from discussions with T. Bar-On, Profs. F.C. Khanna and P.A. Mello.\\

\end{document}